\documentclass[english]{revtex4}
\usepackage{mathptmx}
\usepackage{helvet}

\usepackage[T1]{fontenc}
\usepackage[latin9]{inputenc}
\usepackage{amstext}
\usepackage{amssymb}
\usepackage{graphicx}

\makeatletter

\DeclareRobustCommand{\greektext}{%
  \fontencoding{LGR}\selectfont\def\encodingdefault{LGR}}
\DeclareRobustCommand{\textgreek}[1]{\leavevmode{%
  \IfFileExists{grtm10.tfm}{}{\fontfamily{cmr}}\greektext #1}}
\DeclareFontEncoding{LGR}{}{}
\DeclareTextSymbol{\~}{LGR}{126}
\newcommand{\lyxmathsym}[1]{\ifmmode\begingroup\def\b@ld{bold}
  \text{\ifx\math@version\b@ld\bfseries\fi#1}\endgroup\else#1\fi}

\@ifundefined{textcolor}{}
{%
 \definecolor{BLACK}{gray}{0}
 \definecolor{WHITE}{gray}{1}
 \definecolor{RED}{rgb}{1,0,0}
 \definecolor{GREEN}{rgb}{0,1,0}
 \definecolor{BLUE}{rgb}{0,0,1}
 \definecolor{CYAN}{cmyk}{1,0,0,0}
 \definecolor{MAGENTA}{cmyk}{0,1,0,0}
 \definecolor{YELLOW}{cmyk}{0,0,1,0}
 }

\makeatother

\usepackage{babel}
\begin{document}

\title{Microwave Spectroscopy of a Cooper-Pair Transistor Coupled to a Lumped-Element
Resonator }

\author{Matthew T. Bell, Lev B. Ioffe, and Michael E. Gershenson}

\address{Department of Physics and Astronomy, Rutgers University, 136 Frelinghuysen
Rd., Piscataway, NJ 08854, USA }
\begin{abstract}
We have studied the microwave response of a single Cooper-pair transistor
(CPT) coupled to a lumped-element microwave resonator. The resonance
frequency of this circuit, $f_{r}$, was measured as a function of
the charge $n_{g}$ induced on the CPT island by the gate electrode,
and the phase difference across the CPT, $\phi_{B}$, which was controlled
by the magnetic flux in the superconducting loop containing the CPT.
The observed $f_{r}(n_{g},\phi_{B})$ dependences reflect the variations
of the CPT Josephson inductance with $n_{g}$ and $\phi_{B}$ as well
as the CPT excitation when the microwaves induce transitions between
different quantum states of the CPT. The results are in excellent
agreement with our simulations based on the numerical diagonalization
of the circuit Hamiltonian. This agreement over the whole range of
$n_{g}$ and $\phi_{B}$ is unexpected, because the relevant energies
vary widely, from 0.1K to 3K. The observed strong dependence $f_{r}(n_{g},\phi_{B})$
near the resonance excitation of the CPT provides a tool for sensitive
charge measurements. 
\end{abstract}
\maketitle

\section{Introduction. }

The Cooper-pair transistor (CPT) is a three-terminal device which
consists of a mesoscopic superconducting island connected to two leads
by two Josephson tunnel junctions (JJs) (see, e.g. \cite{Averin1991,Aumentado2010}
and references therein). The behavior of this device is controlled
by two energies: the charging energy per junction, $E_{C}\text{\ensuremath{\equiv}}e^{2}/2C_{J}$
( $C_{J}$ is the capacitance of a single tunnel junction), and the
Josephson coupling energy $E_{J}$. The energies $E_{C}$ and $E_{J}$
could be made of the same order of magnitude by reducing the tunnel
junction in-plane dimensions (typically, down to $100-200nm$ for
$Al-AlO_{x}-Al$ junctions). The energies of quantum states of the
CPT are $2e$ periodic in a continuous charge $n_{g}=C_{g}V_{g}/e$
induced on the island by a capacitively coupled gate electrode. Here
$C_{g}$ is the capacitance of the capacitor formed by the island
and the gate electrode, $V_{g}$ is the voltage applied to this capacitor.
The sensitivity of the CPT characteristics to the induced charge makes
this device a very sensitive electrometer which, in particular, can
operate in a low-dissipation dispersive mode \cite{Delsing2005,Aumentado2006}.
The interplay between the Josephson effect and Coulomb blockade leads
to a quantum superposition of charge states in the CPT, which forms
the basis for quantum computing with superconducting charge qubits
\cite{Nakamura1999,Vion2002,Wallraff2005}. Since the first demonstration
of the coherent superposition of states in the CPT more than a decade
ago, the CPT has been used as a test bed for many novel experimental
techniques employed in the research on superconducting qubits. 

The microwave experiments with CPTs can be broken down into two main
categories. In the first type of measurements, the CPT remains in
its ground state because of a large mismatch between the probe signal
frequency and the excitation frequencies of the CPT. During this adiabatic
operation, the CPT can be described by its effective microwave impedance.
This impedance, depending on the parameters of the Josephson junctions
and the coupling of the CPT to the readout circuit, could be predominantly
inductive (the Josephson inductance, the second derivative of the
CPT energy in phase \cite{Hakonen2009}) or capacitive (the quantum
capacitance, the second derivative of the CPT energy in charge \cite{Widom1984,Averin1985,Hakonen2005}).
Note that if the CPT is coupled to a resonator and their levels are
close in energy, the entanglement of the CPT and resonator states
affects the impedance of this circuit even if the microwaves do not
induce transitions between the states. In the latter case, the impedance-based
description of the CPT is insufficient, and the solution of the quantum
Hamiltonian of the system «CPT + read-out circuit» is required. In
the second type of measurements, the microwaves induce transitions
between different quantum states of the CPT. This, in particular,
enables the preparation and manipulation of coherent superpositions
of the ground and excited states in the quantum-computing-related
applications of the CPT. 

In this paper, we present the microwave spectroscopic study of a CPT
which probes both the ground state and excited states of the CPT over
wide ranges of the charge $n_{g}$ and the phase difference across
the CPT, $\phi_{B}$. The phase was controlled by the magnetic flux
in the superconducting loop containing the CPT. The CPT microwave
response was analyzed by measuring the resonance frequency $f_{r}$
of a lumped-element microwave resonator coupled to a CPT. A relatively
high quality factor of this circuit allowed us to perform the measurements
in the low-power regime with an average number of photons in the resonator
less than one. When the detuning between the microwave frequency and
the excitation frequencies of the CPT was large, the dependence $f_{r}(n_{g},\phi_{B})$
mostly reflected the variations of the CPT Josephson inductance with
$n_{g}$ and $\phi_{B}$. On the other hand, an avoided crossing of
the CPT and resonator levels was clearly observed when the CPT excitation
frequency was tuned to the resonator frequency by varying $n_{g}$
and $\phi_{B}$. The strong $n_{g}$-dependence of the circuit response
in this regime provides a tool for sensitive charge measurements.
The overall dependence $f_{r}(n_{g},\phi_{B})$ is in excellent agreement
with the simulations based on the numerical diagonalization of the
circuit Hamiltonian. This agreement provides a stepping stone for
the understanding of more complicated superconducting circuits intended
for quantum computing, including multi-junction circuits envisioned
as protected superconducting qubits \cite{Doucot2007,Gershenson2009}. 

The paper is organized as follows. In Section \ref{sec:Device-Fabrication-and},
we describe the samples and measurement techniques. The details of
numerical simulations of this circuit are provided in Section \ref{sec:Model-Hamiltonian-and}.
The experimental results are discussed and compared with numerical
simulations in Section \ref{sec:Experimental-Results-and}.

\section{Device Fabrication and Measuring Techniques.\label{sec:Device-Fabrication-and}}

\subsection{Circuit Design}

\begin{figure}
\includegraphics[width=5in]{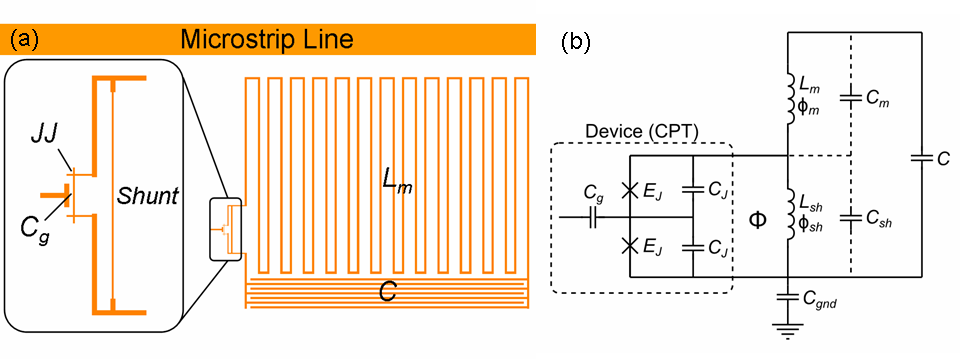}

\caption{Schematics of the {}``CPT+LC resonator'' circuit. (a) The on-chip
circuit layout. The superconducting shunting wire ({}``shunt'')
serves two purposes: it reduces the coupling of the CPT to the LC
resonator (and, thus, reduces external noises), and it forms, in combination
with the CPT, a superconducting loop. The magnetic flux in the loop
controls the phase difference across the CPT. (b) The circuit diagram
used for modeling in Section 3.}

\label{Fig:DeviceSchematics}
\end{figure}

The schematics of the tested circuit is shown in Fig.\ref{Fig:DeviceSchematics}.
The CPT is inductively coupled via a narrow Al wire ({}``shunt'')
to a lumped-element LC resonator. The LC resonator consists of a meandered
2-\textgreek{m}m-wide Al wire with $L=5$ nH and an interdigitated
capacitor (2-\textgreek{m}m-wide fingers with 2 \textgreek{m}m spacing
between them) with $C=100$ fF. It was designed to be strongly coupled
to the microstrip line in order to increase the signal-to-noise ratio
in the low-microwave-power measurements with the average photon occupancy
of the resonator $n_{ph}<1$. The typical values of the internal and
loaded quality factors for these resonators (not coupled to the CPT)
were 50,000 and 20,000, respectively. High Q values enable sensitive
measurements of small changes in the microwave impedance of the tested
device induced by the variations of $n_{g}$ and $\phi_{B}$. Outside
of its bandwidth, the resonator efficiently decouples the CPT from
external noises. An additional protection of the CPT from external
noises is provided by the shunt: the kinetic inductance of this superconducting
wire, $L_{shunt}=0.5$ nH, is more than 10 times smaller than the
effective Josephson inductance of the CPT, which significantly reduces
the phase fluctuations across the CPT. The LC resonator was inductively
coupled to a 2-port Al microstrip line with a 50 \textgreek{W} wave
impedance. The gate electrode of the CPT is coupled to the central
island of the CPT through a capacitor $C_{g}=0.2$ fF. 

The sample was mounted inside an $rf$-tight copper box which provided
the ground plane for the microstrip line and LC resonator. This box
was housed inside another $rf$-tight copper box in order to attenuate
stray photons which may originate from warmer stages of the cryostat
\cite{Barends2011}. This nested-box construction was anchored to
the mixing chamber of a cryogen-free dilution refrigerator at a base
temperature of 20 mK.

\subsection{Device Fabrication}

The Cooper-pair transistor, the lumped-element LC resonator, and the
microstrip line were fabricated within the same vacuum cycle using
multi-angle electron-beam deposition of Al films through a nanoscale
lift-off mask. To minimize the spread of the junction parameters,
we have adopted the so-called \textquotedblleft{}Manhattan-pattern\textquotedblright{}
bi-layer lift-off mask formed by a 400-nm-thick e-beam resist (the
top layer) and 50-nm-thick copolymer (the bottom layer) (see, e.g.,
\cite{Dolan1977,Born2001} and references therein). In this technique,
Josephson junctions are formed between the aluminum strips of a well-controlled
width overlapping at a right angle. After depositing the photoresist
on an undoped Si substrate and its patterning with e-beam lithography,
the sample was placed in an ozone asher to remove any traces of the
photoresist residue. This step is crucial for reducing the spread
in junction parameters. The substrate was then placed in an oil-free
high-vacuum chamber with a base pressure of $5\times10^{-9}$ Torr.
The axis of the rotatable substrate holder forms an angle of $45^{0}$
with the direction of e-gun deposition of Al. During the first Al
deposition, the substrate holder was positioned such that only the
\textquotedblleft{}avenues\textquotedblright{} were covered with metal;
no metal was deposited in the \textquotedblleft{}streets\textquotedblright{}
because the mask thickness is greater than the width of the \textquotedblleft{}streets\textquotedblright{}.
Since the mask thickness is $0.45\mu m$, this technique is suitable
for the fabrication of JJs with lateral dimensions up to $0.3\times0.3\mu m^{2}$.
The thickness of this first Al film, which forms the central island
of the CPT, was 20 nm. Without breaking vacuum, the surface of the
bottom electrodes was oxidized at 100 mTorr of dry oxygen for 5 minutes.
After evacuating oxygen, the substrate holder was rotated by $90^{0}$,
and 60-nm-thick top electrodes were deposited along the \textquotedblleft{}streets\textquotedblright{}
(no aluminum is deposited in the \textquotedblleft{}avenues\textquotedblright{}
at this stage). The central island of the Cooper-pair transistor was
always deposited during the first Al deposition, and its thickness
was smaller than that of the leads; this is important for preventing
the quasiparticle poisoning \cite{Aumentado2004}. Finally, the sample
was removed from the vacuum chamber and the lift-off mask was dissolved
in the resist remover. The spread of the resistances for the nominally
identical JJs with an area of $0.15\times0.15\mu m^{2}$ did not exceed
10\%. More than 10 devices with $E_{J}/E_{C}=1.5-3$ have been studied
and the results have been successfully fitted with the numerical simulations;
below we discuss two representative samples.

\subsection{Measurement Technique }

\begin{figure}
\includegraphics[width=4in]{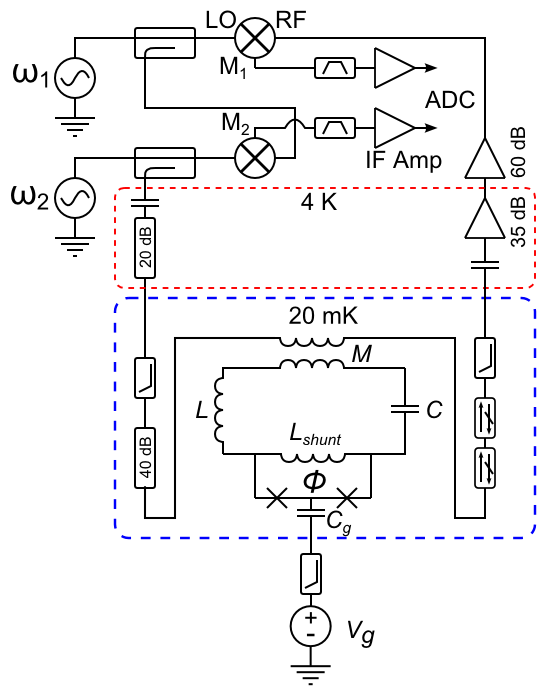}

\caption{Simplified circuit diagram of the measurement setup. The microwave
signal at $\omega_{2}$ transmitted through the microstrip line coupled
to the \textquotedblleft{}LC resonator + CPT\textquotedblright{} circuit
is amplified, mixed down to an intermediate frequency $\omega_{1}-\omega_{2}$,
and digitized by a fast digitizer (ADC). The second channel of the
ADC is used to digitize the signal from an additional mixer (M2),
which provided the reference phase $\varphi_{0}$ (see the text).
The gate voltage $V_{g}$ is applied to the capacitor $C_{g}$ using
a heavily filtered DC line. }

\label{Fig:MeasurementSchematics}
\end{figure}

The microwave response of the coupled system \textquotedblleft{} CPT
+ LC resonator\textquotedblright{} was probed by measuring the amplitude
and phase of the microwaves traveling along a microstrip line coupled
to the resonator. Figure \ref{Fig:MeasurementSchematics} shows a
simplified schematic of the microwave circuit, which is similar to
the one used in work \cite{Manucharyan2009}. The cold attenuators
and low-pass filters in the input microwave line prevented leakage
of thermal radiation into the resonator. On the output line, a combination
of low-pass filters and two cryogenic Pamtech isolators (\textasciitilde{}18
dB isolation between 3 and 12 GHz) anchored to the mixing chamber
were used to attenuate the 5 K noise from the cryogenic amplifier.
The DC line for the gate voltage control was heavily filtered with
a combination of room temperature LC and low temperature RC filters,
followed by a stainless steel powder filter, and a 1:1000 voltage
divider. 

The probe signal at frequency $\lyxmathsym{\textgreek{w}}_{2}$, generated
by a microwave synthesizer (Anritsu MG3694B), was coupled to the cryostat
input line through a 16 dB coupler. This signal, after passing the
sample, was amplified by a cryogenic HEMT amplifier (Caltech CITCRYO
1-12, 35 dB gain between 1 and 12 GHz) and two 30 dB room-temperature
amplifiers. The amplified signal was mixed by mixer M1 with the local
oscillator signal at frequency $\omega_{1}$, generated by another
synthesizer (Gigatronics 910). The intermediate-frequency signal $a(t)=a\sin(\lyxmathsym{\textgreek{W}}t+\varphi)+noise(t)$
at $\lyxmathsym{\textgreek{W}}\equiv(\lyxmathsym{\textgreek{w}}_{1}-\lyxmathsym{\textgreek{w}}_{2})/2\lyxmathsym{\textgreek{p}}=30$
MHz was digitized by a $1$ GS/s digitizing card (AlazarTech ATS9870).
The signal was digitally multiplied by $\sin(\lyxmathsym{\textgreek{W}}t)$
and $\cos(\lyxmathsym{\textgreek{W}}t)$, averaged over many (typically,
$10^{6}$) periods, and its amplitude $a$ (proportional to the microwave
amplitude $S_{21}$) and phase $\varphi$ were extracted as $a=\sqrt{\left\langle a^{2}(t)\sin^{2}\lyxmathsym{\textgreek{W}}t+a^{2}(t)\cos^{2}\lyxmathsym{\textgreek{W}}t\right\rangle }$
and $\varphi=\arctan\left(\left\langle a^{2}(t)\sin^{2}\Omega t\right\rangle /\left\langle a^{2}(t)\cos^{2}\Omega t\right\rangle \right)$,
respectively (here $\left\langle ...\right\rangle $ stands for the
time averaging over integer number of periods). The value of $\varphi$
randomly changes when both $\omega_{1}$ and $\omega_{2}$ are varied.
To eliminate these random variations, we have also measured the phase
$\varphi_{0}$ of the reference signal provided by mixer M2 and digitized
by the second channel of the ADC. The phase difference $\varphi-\varphi_{0}$,
being dependent at fixed $n_{g}$ and $\phi_{B}$ only on the electric
length difference between the microwave lines inside and outside of
the cryostat, is immune to the phase jitter between the two synthesizers.
The low noise of this setup allowed us to perform measurements at
microwave excitation level down to -140 dBm which corresponded to
sub-single-photon population of the tank circuit. 

\section{Model Hamiltonian and Numerical Simulations\label{sec:Model-Hamiltonian-and}}

We begin with the discussion of the theoretical model of a more general
circuit which contains an arbitrary Josephson device coupled via a
superconducting {}``shunt'' to a microwave resonator. The only limitation
on the device parameters is that all characteristic energies of the
device are much smaller than the effective inductive energies of all
superconducting wires in the circuit, $E_{L_{i}}=\hbar^{2}/(2e)^{2}L_{i}$.
The resonance frequency of the circuit might be of the same order
or even very close to the device excitation energies, which would
lead to the level repulsion. To simplify the notations, we shall use
below the units $\hbar=2e=1$ (e.g., in these units $V=d\phi/dt$)
and restore the physical units at the end, where we apply this model
to the specific case of a device that consists of two Josephson junctions
and one superconducting island, i.e. the CPT. 

The generalized circuit shown in Fig. 1b includes two loops: the long
meandering wire, the shunt, and a large capacitor $C$ form one loop
(referred below as the {}``resonator'' loop), and the device and
the shunt form another loop (referred as the {}``device'' loop).
The inductance of the meander (shunt), and the phase difference across
this element are $L_{m}$ ($L_{sh}$) and $\phi_{m}$ ($\phi_{sh}$),
respectively. The difference between the device phase $\phi_{0}$
and the shunt phase $\phi_{sh}$ is due to the time-independent magnetic
flux $\Phi$ in the device loop: $\phi_{0}-\phi_{sh}=\phi_{B}$, where
$\phi_{B}=2\pi\Phi/\Phi_{0}$, $\Phi_{0}$ is the flux quantum. The
voltage drops across the device and the shunt are equal: $V_{sh}=V_{D}$.
The whole circuit is described by the Lagrangian
\begin{eqnarray}
\mathit{\mathcal{L}} & = & T_{sh}(V_{sh})+T_{m}(V_{m})+\frac{C}{2}(V_{sh}+V_{m})^{2}-\frac{1}{2}E_{sh}\phi_{sh}^{2}-\frac{1}{2}E_{m}\phi_{m}{}^{2}\label{eq:L_tot}\\
 & + & \mathcal{L}_{D}(\phi_{0},V_{0}).\nonumber 
\end{eqnarray}
Here $T_{m}(V)$ ($T_{sh}(V)$) is the time-dependent part (i.e. the
kinetic energy) of the response of the meander (shunt), and $\mathcal{L}_{D}(\phi_{0},V_{0})$
is the device Lagrangian which also depends on the internal degrees
of freedom (phases) of the device. In the BCS theory, the response
of a superconducting wire with the static energy $E_{L}$ has a scale
of $E_{L}$ at all frequencies $\omega\lesssim\Delta$; it is a function
of the dimensionless parameter $V/\Delta$: $T=E_{L}f(V/\Delta)=(1/16)(V/\Delta)^{2}+O((V/\Delta)^{4})$
where $\Delta$ is the superconducting gap. This equation implies
that at low frequencies, the wire impedance acquires, in addition
to the kinetic inductance, a small capacitive component $C'=E_{L}/(8\Delta^{2})$
(these capacitances are shown in Fig.\,1b by dashed lines). 

We shall assume that the device Lagrangian is given by the sum of
the Josephson and electrostatic energies:
\begin{equation}
\mathcal{L}_{D}(\phi_{D},V_{D})=\frac{1}{2}\sum_{i,j}C_{ij}V_{i}V_{j}+\sum_{i,j}J_{ij}\cos(\phi_{i}-\phi_{j}-\Phi_{ij})\label{eq:L_D}
\end{equation}
Here phases $\phi_{i}$ and corresponding potentials $V_{i}$ describe
both the internal degrees of freedom of the device and the shunt phase. 

Because the potential energy of the resonator loop is much greater
than that of the device, the effect of the device on this loop can
be treated as a small perturbation. In the absence of the device,
the resonator loop has two modes: the harmonic oscillation of the
total phase $\phi_{sh}+\phi_{m}$ with frequency $\omega_{0}=\sqrt{1/L'C}$
where $L'=E_{m}^{-1}+E_{sh}^{-1}$, $E_{sh}\phi_{sh}=E_{m}\phi_{m}$,
and the orthogonal mode with $\phi_{m}+\phi_{sh}\approx0$. Because
the large capacitance $C$ does not participate in the second mode,
the frequency of this mode is determined by the superconducting gap,
the only energy scale in this case. We shall assume that $\Delta\gg\omega_{0}$
so both real and virtual excitations of this mode can be neglected
as well as the contribution of the capacitances $C'_{sh}$ and $C'_{m}$
to the effective capacitance $C$ of the first mode. Note, however,
that the virtual processes involving the second mode are small only
in $\omega_{0}/\Delta$ and might not be completely negligible in
a realistic situation. If these processes are neglected, there is
only one relevant degree of freedom, the phase across the device $\phi_{0}$.
The effective Lagrangian is reduced to
\begin{equation}
\mathcal{L}_{eff}=\frac{C_{L}}{2}V_{0}^{2}-\frac{1}{2}E_{L}(\phi_{0}-\phi_{B}){}^{2}+\mathcal{L}_{D}(\phi_{0},V_{0})\label{eq:L_eff}
\end{equation}
 with $C_{L}=C(1+E_{sh}/E_{m})^{2}$ and 

\begin{equation}
E_{L}=E_{sh}(1+E_{sh}/E_{m}).\label{eq:E_L}
\end{equation}

Fluctuations of the phase $\phi_{0}$ in the low-energy states of
the oscillator mode are very small: 
\begin{eqnarray*}
\left\langle (\phi_{0}-\phi_{B}){}^{2}\right\rangle  & = & A^{2}(2n+1)\ll1\\
A^{2} & = & \frac{\omega_{0}}{2E_{L}}
\end{eqnarray*}
where $n$ is the quantum number of the oscillator states. This allows
one to replace the solution of the full problem by the solution of
the simplified model in which we expand the interaction term in small
phase fluctuations. 

It will be more convenient to use the Hamiltonian formalism in which
the conjugated degrees of freedom are phases and charges. The total
Hamiltonian is the sum of three parts, the Hamiltonians of the resonator
($H_{R}$), device ($H_{D}),$ and interaction between them ($H_{int}$)
: 
\begin{eqnarray}
H_{R} & = & \frac{\omega_{0}^{2}}{2E_{L}}q_{0}^{2}+\frac{1}{2}E_{L}(\phi_{0}-\phi_{B}){}^{2}\label{eq:H_R}\\
H_{int} & = & C_{L}^{-1}\sum_{i,j>0}q_{0}C_{0j}C_{ji}^{-1}(\mathbf{q}_{ji}-n_{i})-\sum_{i}J_{i0}\cos(\phi_{0}-\phi_{j}-\Phi_{0i})\label{eq:H_int}\\
H_{D} & = & \frac{1}{2}\sum_{ij>0}(\mathbf{q}_{\mathbf{i}}-n_{i})C_{ij}^{-1}(\mathbf{q}_{j}-n_{j})-\frac{1}{2}\sum_{ij>0}J_{ij}\cos(\phi_{i}-\phi_{j}-\Phi_{ij})\label{eq:H_D}
\end{eqnarray}
The coupling to the inductor charge fluctuations contains the inverse
of the capacitance matrix ($\sim C_{L}^{-1}$) and thus is very small.
Thus, even though the charge fluctuations across the shunt are not
small:
\[
\left\langle q_{0}{}^{2}\right\rangle =\frac{1}{4A^{2}}(2n+1)\gg1
\]
their effect on the coupling can be treated perturbatively. In the
leading order in the interaction, we need to keep only two types of
terms. The first type is quadratic in phase $\phi_{0}$ and diagonal
in the basis of resonator states. The second type is linear in $\phi_{0}$
and off diagonal in this basis. The quadratic terms in the inductor
charge are absent, so the charge coupling appears only due to the
off diagonal terms that are linear in $q_{0}$. The Hamiltonian equivalent
to (\ref{eq:L_eff}) becomes
\begin{equation}
H_{eff}=(\omega_{0}+2A^{2}\Xi)(a^{\dagger}a+1/2)+(A\mathbf{J}+\frac{1}{4A}\mathbf{Q})a+h.c.\label{eq:H_eff}
\end{equation}
 Here $a\dagger$ ($a$) is the creation (annihilation) operator for
the harmonic oscillations of the circuit, $\mathbf{J},$ $\mathbf{Q}$
and $\Xi$ are operators acting on the device which forms are obtained
by expanding the interaction Hamiltonian 
\begin{eqnarray}
\mathbf{J} & = & \left.\frac{dL}{d\phi_{0}}\right|_{\phi_{0}=0}=-\sum_{i}J_{i0}\sin(\phi_{j}+\Phi_{0i})\label{eq:J}\\
\mathbf{Q} & =C_{L}^{-1} & \sum_{i,j>0}C_{0j}C_{ji}^{-1}(\mathbf{q}_{ji}-n_{i})\label{eq:Q}\\
\Xi & = & \frac{1}{2}\left.\frac{d^{2}L}{d\phi_{0}^{2}}\right|_{\phi_{0}=0}=\frac{1}{2}\sum_{i}J_{i0}\cos(\phi_{j}+\Phi_{0i})\label{eq:Xi}
\end{eqnarray}

We now estimate the scale of the frequency deviations induced by these
perturbations. In the natural units of the resonator frequency $\omega_{0}$,
the scales of the perturbing operators are $A\mathbf{J}/\omega_{0}\sim E_{J}/\sqrt{\omega_{0}E_{L}}$,
$A^{-1}\mathbf{Q}/\omega_{0}\sim\sqrt{\omega_{0}/E_{L}}$ and $A^{2}\Xi/\omega_{0}\sim E_{J}/E_{L}$.
The operator $\Xi$ is diagonal in the oscillator states, so it directly
results in the frequency shift $\delta\omega_{\Xi}/\omega_{0}\sim E_{J}/E_{L}.$
Because the non-diagonal elements affect the level of the resonator
only in the second order of the perturbation theory, the effect of
the $\mathbf{J}$ and $\mathbf{Q}$ operators depends on the gap between
the levels in the combined resonator/device circuit. Far away from
the full frustration and charge degeneracy point ($\phi_{B}=\pi$,
$n_{g}=0.5$), the device is characterized by large $E_{J}\gg\omega_{0}$
and the energy levels are separated by large gaps, so the smallest
gap is due to the resonator: $\delta E=\omega_{0}$. In this case
the frequency shifts are $\delta\omega_{J}/\omega_{0}\sim E_{J}^{2}/(\omega_{0}E_{L})$
and $\delta\omega_{Q}/\omega_{0}\sim\omega_{0}/E_{L}$ respectively,
which implies $\delta\omega_{J}\gg\delta\omega_{\Xi}\gg\delta\omega_{Q}$.
The effect induced by the phase and charge coupling grows when the
gap between the levels coupled by these operators becomes small, but
the phase coupling remains larger than the charge coupling for the
devices with $E_{J}\gg\omega_{0}.$ This increase of the frequency
shift occurs, for instance, when the device level crosses the first
resonator level. 

We now write down the explicit equations for the Cooper pair box.
In this case, the internal degrees of freedom are limited to one phase
(and the conjugated charge). Assuming equal capacitances and Josephson
energies of the CPT junctions, we have 
\begin{eqnarray}
H_{CPT} & = & 4E_{c}(q_{1}-n_{g})^{2}-E_{J}\left[\cos(\phi_{1})+\cos(\phi_{1}+\phi_{B})\right]\label{eq:H_CPT}\\
\mathbf{J} & = & -E_{J}\sin(\phi_{1}+\phi_{B})\label{eq:J_CPT}\\
\mathbf{Q} & = & \frac{2C}{C_{L}}E_{c}(q_{1}-n_{g})\label{eq:Q_CPT}\\
\Xi & = & \frac{1}{2}E_{J}\cos(\phi_{1}+\phi_{B})\label{eq:Xi_CPT}
\end{eqnarray}
Here we restored the physical energy units $E_{c}=e^{2}/2C_{J}$.
For practical computations it is sufficient to retain the first few
levels of the resonator ($a^{\dagger}a\leq n_{max}=3)$ and some number,
$n_{Q}$, of the charging states. The Hamiltonian (\ref{eq:H_eff})
becomes $3n_{Q}\times3n_{Q}$ matrix. Because the wave function of
the charge decreases exponentially at large charges, $\Psi(q)\sim\exp(-\sqrt{E_{c}/E_{J}}q^{2})$,
it is sufficient to consider $n_{Q}\sim10$ for accurate computations.
The straightforward numerical digonalization of the Hamiltonian (\ref{eq:H_eff})
leads to the theoretical predictions that can be compared with the
data. 

\begin{figure}
\includegraphics[width=5in]{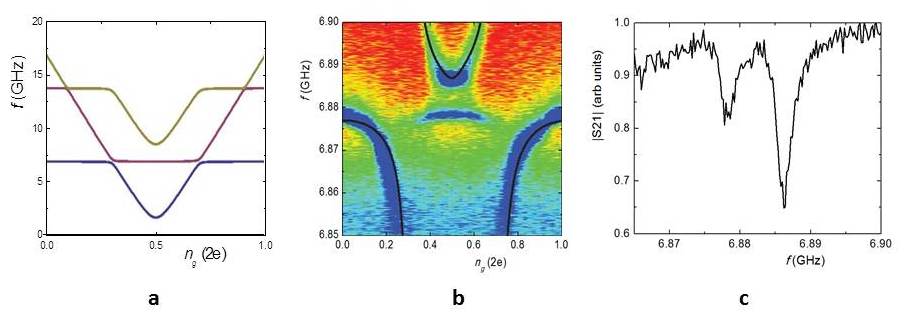}

\caption{(color online) Panel $\mathbf{a}$ shows the frequencies of three
lowest-energy levels for the coupled system {}``CPT + LC resonator''
plotted as a function of $n_{g}$ at a constant phase $\phi_{B}=0.97\pi$.
The model parameters are $E_{J}=h\times17.2\,\mbox{GHz}$, $E_{C}=h\times8.6\,\mbox{GHz}$
and $E_{L}=h\times5650\,\mbox{GHz}$. In this very frustrated regime,
the lowest level of the CPT crosses the lowest level of the resonator
with approaching the charge degeneracy point $n_{g}=0.5$, which results
in the avoided level crossing. Panel $\mathbf{b}$ shows the blow-up
of the theoretical curves in the region of avoided level crossing
(black curves), together with the dependence of the color-coded microwave
amplitude $S_{21}$ on the microwave frequency $f$ and $n_{g}$ measured
for one of the tested CPTs at $\phi_{B}=0.97\pi$. The minimum of
the microwave amplitude corresponds to the resonance frequency. The
dependence of $S_{21}(f)$ at the charge degeneracy point $n_{g}=0.5$
is shown in Panel $\mathbf{c}.$}

\label{Fig:GlobalLevelStructure}
\end{figure}
Our experimental situation corresponds to $E_{c}\sim2\hslash\omega_{0}$
and $E_{J}\sim4\hbar\omega_{0}$. In the absence of frustrations,
the frequency of the lowest CPT level is very high: $\omega_{p}=\sqrt{8E_{c}E_{J}}\sim10\omega_{0}$.
The frequency of the lowest CPT level decreases as the magnetic field
frustrates the Josephson coupling and/or with approaching the charge
degeneracy ($n_{g}=0.5)$. Figure \ref{Fig:GlobalLevelStructure}a
shows three low-energy levels of the system {}``CPT+resonator''
with the parameters typical for our experiment. Note that for the
studied circuits, only the combined effect of flux- and charge-induced
frustrations brings the frequency of the first device level below
that of the resonator, otherwise the device resonance frequency significantly
exceeds that of the resonator even at full flux frustration (e.g.
$\omega=2E_{c}>4\omega_{0}$ at $n_{g}=0$).

\section{Experimental Results and Discussion\label{sec:Experimental-Results-and}}

\begin{figure}
\includegraphics[width=5in]{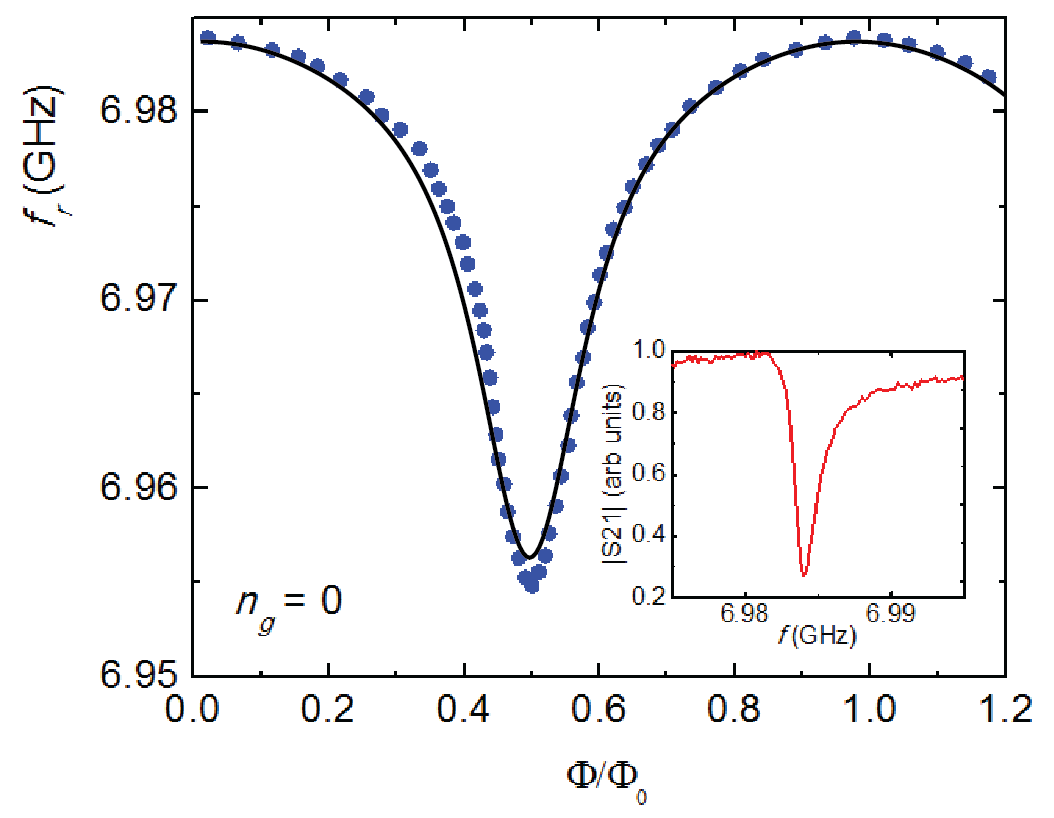}

\caption{Dependence of the resonance frequency $f_{r}(n_{g}=0,\phi_{B})$ on
the magnetic flux $\Phi$, which controls the phase difference across
the CPT, $\phi_{B}=2\pi\Phi/\Phi_{0}$. The solid curve shows the
numerical simulation with the fitting parameters discussed in the
text. The inset shows the dependence of the microwave amplitude $S_{21}$
on the frequency near the resonance at $\phi_{B}=0$ and $n_{g}=0$.}

\label{Fig:Resonance}
\end{figure}

Below we present the measurements of the amplitude $S_{21}$ of the
transmitted microwaves (unless otherwise specified) at the base temperature
$T=20$ mK. Most of the data (with the exception of the data in Fig.3b,c)
are shown for only one representative device. The resonant dependence
of $S_{21}$ on the microwave frequency $f$ , measured for this device
at $\phi_{B}=0$ and $n_{g}=0$, is shown in the inset to Fig. \ref{Fig:Resonance}.
The resonance frequency depends periodically on $n_{g}$ and $\phi_{B}$;
for example, the dependence $f_{r}(n_{g},\phi_{B})$ measured at $n_{g}=0$
is shown in Figure \ref{Fig:Resonance}. The period in charge is $\Delta n_{g}=2e$
at the base temperature (see Fig.5); it changes from $2e$ to $e$
at higher temperatures (>300 mK) due to the presence of thermally
excited quasiparticles. Note that the total time of acquisition for
the data shown in Fig.3b was approximately 20 minutes; over longer
time intervals, the periodicity of $f_{r}(n_{g},\phi_{B})$ might
be disrupted by the motion of non-equilibrium quasiparticles to/from
the CPT island (the so-called {}``quasiparticle poisoning'') \cite{Aumentado2004}
or other types of charge fluctuations \cite{Faoro2006}. The high
stability of the charge on the CPT island indicates that (a) the combination
of a larger superconducting gap of the CPT island and its relatively
large charging energy protects the CPT from quasiparticle poisoning,
and (b) the double-wall $\mathit{rf}$-tight sample box shields the
device from stray high-energy photons. The microwave photon energy
$E_{ph}\thickapprox h\times7$ GHz is insufficiently large to excite
the CPT at $n_{g}=0$: indeed, according to our simulations, the lowest
excitation frequency for this device exceeds 30 GHz even at full flux
frustration ($\Phi/\Phi_{0}=0.5$). In this case the variation of
the resonance frequency $f_{r}$ with magnetic flux reflects the $\phi_{B}$-dependence
of the CPT impedance in its ground state. 

\begin{figure}
\includegraphics[width=5in]{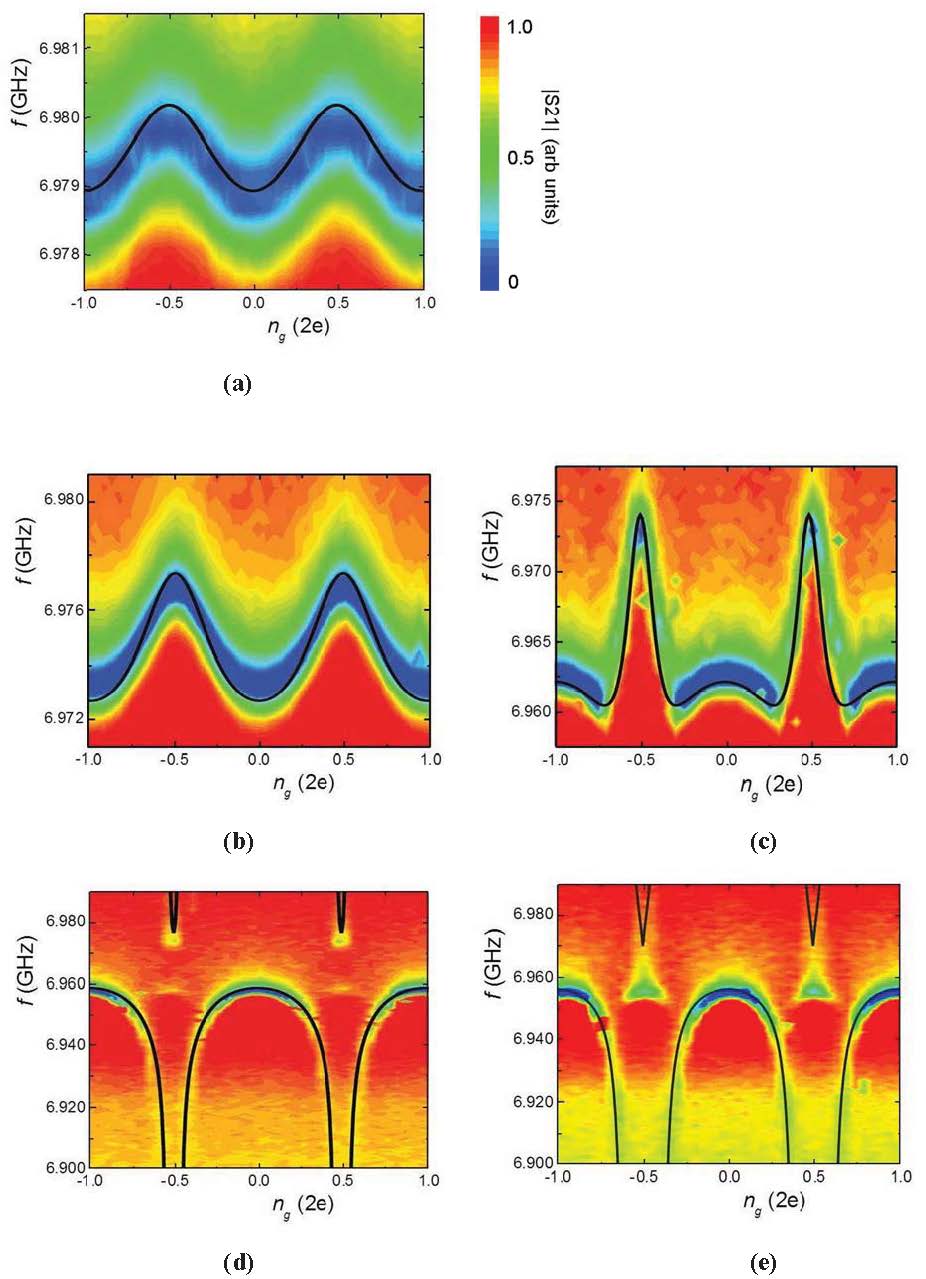}

\caption{The color-coded plots of the microwave amplitude \mbox{l}$S_{21}\text{\mbox{l}}$
versus the microwave frequency $f$ and the charge $n_{g}$ induced
on the CPT island by the gate voltage. The phase difference across
the CPT was controlled by the magnetic flux $\Phi$ in the device
loop: (a) $\Phi/\Phi_{0}=0.58$ , (b) $\Phi/\Phi_{0}=0.75$, (c) $\Phi/\Phi_{0}=0.90$,
(d) $\Phi/\Phi_{0}=0.94$, (e) $\Phi/\Phi_{0}=1$. The solid curves
show the numerical simulations with the fitting parameters discussed
in the text. }

\label{Fig:2Dplots}
\end{figure}

The dependences of the resonance frequency on $n_{g}$ are illustrated
by Figs. \ref{Fig:2Dplots}(a-e), where the color-coded microwave
amplitude $S_{21}$ is plotted versus $f$ and $n_{g}$ for several
values of the magnetic flux in the device loop. The black curves in
Fig. \ref{Fig:2Dplots} show the results of fitting the experimental
data with our numerical simulations. All these curves were generated
with the same set of fitting parameters: $E_{C}=h\times16$ GHz, $E_{J}=h\times32$
GHz and $E_{L}=h\times5720$ GHz (note that not only the amplitude
of the resonance frequency modulation, but also the absolute values
of $f_{r}$ are pre-determined by these parameters). The fitting procedure
is very sensitive to the choice of these parameters: we believe that
they are determined with an accuracy better than 10\%. The extracted
charging energy coincides (within 5\% accuracy) with an estimate of
$E_{C}$ based on the junction area, the specific geometrical capacitance
for Al tunnel junctions ($50$ fF/$\mu m^{2}$, see e.g,. \cite{Desling1994}),
and the electronic capacitance of Josephson junctions, $C_{e}=3/16(R_{Q}/R)e^{2}/\Delta$
(0.3 fF at $R=3$ k\textgreek{W}) \cite{Larkin1983,Ambegaokar1984}.
The Josephson energy estimated on the basis of the Ambegaokar-Baratoff
relationship \cite{Tinkham} using the normal-state resistance of
a test junction deposited on the same chip is \textasciitilde{}40\%
greater than the fit value of $E_{J}$. 

Generally, one expects that Josephson circuits can be accurately described
by the Hamiltonian consisting of Josephson and charging energies (cf.
Eq.\ref{eq:L_D}) only if all energy scales are smaller than $\Delta$.
Away from full frustration, the energy of the CPT excited state is
of the order of Josephson plasma frequency $\approx3.2$K, which is
comparable to $\Delta$. Thus, the excellent agreement between the
experimental data and numerical modeling, observed over the whole
range of $n_{g}$ and $\phi_{B}$, is quite surprising.

It is worth noting that the circuit modeling based on the numerical
diagonalization of the circuit Hamiltonian is essential for fitting
the data for devices with $E_{J}/E_{C}\sim2$. For example, the analytical
solution for the Josephson inductance in the CPT ground state, calculated
within the two-level approximation (Eq. 4 in Ref. \cite{Hakonen2009}),
overestimates the amplitude of the $f_{r}\left(n_{g}\right)$ dependence
at small flux frustrations by almost an order of magnitude. The latter
solution provides more accurate fitting of the experiment at larger
frustrations ($\phi_{B}\sim(0.8-9)\pi$), but becomes inadequate again
at $\phi_{B}>0.94\pi$ when the avoided level crossing is observed.

The evolution of these dependences reflects the modification of the
CPT spectrum with $n_{g}$ and $\phi_{B}$. For a small phase difference
$\phi_{B}$ (Figs.\ref{Fig:2Dplots}a-b), the lowest CPT excitation
frequency well exceeds the microwave frequency (which is close to
the resonance frequency of the resonator), and the CPT remains in
its ground state for all $n_{g}$ including the charge degeneracy
point ($n_{g}=0.5$). In this regime, the dependences $S_{21}(f,n_{g})$
mostly reflect the variations of the CPT impedance with $n_{g}$ in
the CPT ground state. At a larger frustration $\phi_{B}=0.9\pi$ (Fig.
\ref{Fig:2Dplots}c), the lowest CPT level approaches the lowest resonator
level $n_{g}=0.5$ (but has not crossed it yet), and the entanglement
of the qubit and resonator states leads to a complicated shape of
the $f_{r}\left(n_{g}\right)$ dependence. Finally, with the further
approach to full frustration ($\phi_{B}\geq0.94\pi$, Figs. \ref{Fig:2Dplots}d,e),
the CPT excitation frequency becomes smaller than the resonance frequency
of the LC resonator, and the shape of the $f_{r}\left(n_{g}\right)$
dependences abruptly changes: they are strongly affected by the avoided
level crossing. The $S_{21}(f,n_{g})$ plots in this regime consist
of two sets of curves. The lower set of curves corresponds to the
lowest energy level of the combined system {}``CPT + resonator''
(this level coincides with the CPT lowest level when approaching the
charge degeneracy points, i.e. far away from the resonance frequency
of the resonator). The upper set of curves corresponds to the first
excited level of the system {}``CPT + resonator'': with approaching
the charge degeneracy point, this level descends from higher energies
to its lowest position at $n_{g}=0.5$. The visibility of the upper
set of curves depends on the proximity between the CPT and resonator
levels. Indeed, if the energy of the CPT resonance at $n_{g}=0.5$
is much lower than the first resonator level, the upper-curve {}``cone''
is very sharp, and the corresponding microwave resonance is smeared
even for small deviations of $n_{g}$ from $0.5$ (this case is illustrated
by Figs. \ref{Fig:2Dplots}d,e). On the other hand, the {}``cone''
becomes broader when the intersecting CPT and resonator levels are
close to one another: in this case illustrated by Fig. 1b, we were
able to follow the upper set of curves over the frequency range of
\textasciitilde{}15 MHz.

For both devices, whose dependences $S_{21}(f,n_{g})$ are shown in
Figs. 3 and \ref{Fig:2Dplots}, we observed a double-resonance structure
at full frustration and charge degeneracy, depicted in Figs. 3b,c
and 5e. The second (low-frequency) resonance appears as a {}``shadow''
of the resonance observed at $n_{g}=0$. The appearance of this resonance,
much weaker than that at $n_{g}=0$, implies that there are fluctuations
of the island offset charge $\pm e$ which are fast at the measuring
time scale \textasciitilde{}0.1 s. These fluctuations change the effective
$n_{g}$ from 0.5 to 0. We attribute these fluctuations to the non-equilibrium
quasiparticles moving between the CPT island and the leads. At $n_{g}=0.5$,
the energy of a quasiparticle on the island exceeds the energy of
quasiparticles in the leads by $\delta\Delta-(1/2)E_{C}$. Here $\delta\Delta$
is the difference between superconducting gaps in the island and the
leads due to the difference in the thicknesses of these Al films;
we estimate $\delta\Delta$ to be \textasciitilde{}$k_{B}\times0.3K$.
In our devices, the probability of these fluctuations is small (the
amplitude of the $n_{g}=0$ resonance is much greater than that of
its {}``shadow'' at $n_{g}=0.5$), which suggests that the quantity
$\delta\Delta-(1/2)E_{C}$ remains positive (albeit small) at $n_{g}=0$. 

The sensitivity of the studied circuit to the charge on the CPT island
is illustrated by Fig.6. Figure 6b shows the time dependence of the
phase $\varphi$ of transmitted microwaves when the microwave frequency
is tuned to the resonance of the {}``CPT + resonator'' circuit.
The noise was measured at full flux frustration ($\phi_{B}=\pi$)
when an avoided crossing between the CPT and resonator levels was
observed, but relatively far from the charge degeneracy point ($n_{g}=0.17$).
The amplitude of the observed telegraph noise corresponds to the charge
fluctuations $\Delta q\approx0.03e$ due to, presumably, coupling
of the CPT island to a single charge fluctuator in its environment.

\begin{figure}
\includegraphics[width=5in]{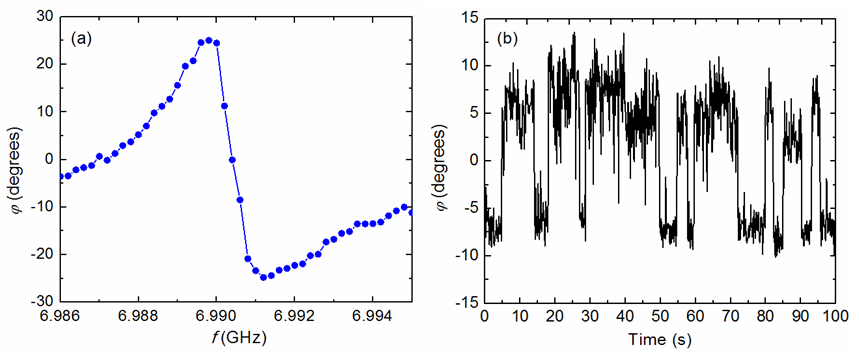}

\caption{Panel a: Dependence of the phase $\varphi$ of the transmitted microwaves
on the microwave frequency near the resonance of the {}``CPT + resonator''
circuit. The CPT is in a strongly frustrated regime ($\phi_{B}=\pi,\; n_{g}=0.17$),
when an avoided crossing between the CPT and resonator levels is observed.
The measurement time for each experimental point is 24 ms. Panel b:
The microwave phase measured with the same averaging time (24ms/point)
at a fixed microwave frequency $f=6.9905$ GHz over a time period
of 100 s. The full range of phases from $-15^{0}$ to $+15^{0}$ corresponds
to the offset charge variation by 0.05e.}

\end{figure}

\section{Conclusions. \label{sec:Conclusions.}}

We have performed a detailed analysis of the microwave response of
Cooper pair transistors with $E_{J}/E_{C}\sim1.5-3$ coupled to a
lumped element resonator as a function of the magnetic flux and the
gate voltage. Away from the full frustration in flux and charge the
levels of the Cooper pair transistor are far away from the resonator
frequency. In this regime the modulation of the resonator frequency
induced by CPT can be described as the effective inductance of the
CPT that adds to the inductance of the resonator. Close to the full
frustration the frequency of the resonator approaches and eventually
crosses the excitation level of the Cooper pair transistor resulting
in a complex pattern of the resonator frequency dependence on the
flux and gate voltage. In all regimes the resonator frequency dependence
is very well described by the results of the numerical diagonalization
of the full Hamiltonian that contains the resonator and CPT. High
sensitivity of the resonator frequency close to the level crossing
provides the tool to measure charge fluctuations in the environment
with high accuracy and short time resolution. 

We would like to thank J. Aumentado and V. Manucharyan for helpful
discussions. We acknowledge the support from the DARPA (under grant
HR0011-09-1-0009), ARO (W911NF-09-1-0395), and NSF (NIRT ECS-0608842). 

\pagebreak{}

\bibliographystyle{apsrev}
\bibliography{QubitSpectroscopy}

\end{document}